\documentclass[aps,prl,twocolumn]{revtex4-2}%
\usepackage{amsfonts}
\usepackage{amssymb}
\usepackage{natbib}
\usepackage{graphicx}
\usepackage{xcolor}
\usepackage{amsfonts}
\usepackage{amsmath}%
\providecommand{\U}[1]{\protect\rule{.1in}{.1in}}

\setcitestyle{numbers,square}
\begin{document}

\title{Theory of Transport in Ferroelectric Capacitors}
\author{Gerrit E. W. Bauer$^{1-5,}$}
\email{g.e.w.bauer@imr.tohoku.ac.jp}
\author{Ryo Iguchi$^{3}$ and Ken-ichi Uchida$^{2-4}$}
\date{\today}

\affiliation{$^{1}$WPI-AIMR, Tohoku
University, Sendai 980-8577, Japan}
\affiliation{$^{2}$Institute for Materials Research,Tohoku
University, Sendai 980-8577, Japan}
\affiliation{$^{3}$National Institute for Materials Science, Tsukuba 305-0047, Japan}
\affiliation{$^{4}$Center for Spintronics Research Network, Tohoku University, Sendai 980-8577, Japan}
\affiliation{$^{5}$Zernike Institute for Advanced Materials, Groningen University, Netherlands}

\begin{abstract}
We present a time-dependent diffusion theory for heat and polarization
transport in a planar ferroelectric capacitor with parameters derived from
a one-dimensional phonon model. We predict steady-state Seebeck and transient
Peltier effects.

\end{abstract}
\maketitle

Magnetism and ferroelectricity are caused by the spontaneous ordering of
magnetic and electric dipoles, respectively, and have much in common
\cite{Spaldin}. Electro- and magnetocaloric phenomena are similar entropic
effects caused by a transient non-equilibrium in the electric and magnetic
dipole ensembles \cite{Crossley}. Transport phenomena attracted more attention
in magnetic structures \cite{Bauer,Back}. In ferroelectrics, a
thermopolarization\ \cite{Marvan,Gurevich,TrepakovTP} and dielectric
Peltier effect \cite{Tagantsev,TrepakovPeltier} have been reported, but the associated theory inappropriately mixes caloric and transport (or caloritronic) phenomena. 

Here we formulate a linear-response  theory of transport in electrically insulating ordered ferroelectrics (FE) for the most basic device, i.e. a slab of a FE between metallic contacts. We solve the diffusion
problem with emphasis on the time response to an abrupt perturbation. A
microscopic theory of polarization and heat currents based on a
one-dimensional phonon model for a displacive FE leads to numerical parameter estimates. The analogies and differences with magnon transport are illuminating \cite{Cornelissen,Flebus}.

\emph{Linear response -} We consider a FE with metal contacts in a planar configuration with
polarization density $p\left(  x\right)$ normal to the interfaces, which interacts with an electric field \(E(x)\) as
\begin{equation}
H=-\int p\left(  x\right)  E\left(  x\right)  dx.
\end{equation}
Temperature ($\partial T)$ and electric field ($\partial E)$ gradients
generate heat $j_{q}$ and polarization $j_{p}$ current densities. In the
linear response regime, \textquotedblleft currents\textquotedblright\ and
\textquotedblleft forces\textquotedblright\ are related by a matrix of
material and device-dependent transport coefficients \cite{Groot,Landau}. The
caloritronic relations can be summarized by a $2\times2$ linear response
matrix%
\begin{equation}
\left(
\begin{array}
[c]{c}%
j_{p}\\
j_{q}%
\end{array}
\right)  =\sigma\left(
\begin{array}
[c]{cc}%
1 & S\\
\Pi & \kappa/\sigma
\end{array}
\right)  \left(
\begin{array}
[c]{c}%
\partial E\\
-\partial T
\end{array}
\right)  , \label{onsager}%
\end{equation}
where $\sigma$ $\left(  \kappa\right)  $ is the polarization (thermal)
conductivity with units m/$\mathrm{\Omega}$ (W/m/K), while $S$ $\left(
\Pi\right)  $ is the ferroelectric Seebeck (Peltier) coefficient with units
V/K/m (V/m). The dissipation rate $\dot{f}=j_{p}\partial E+j_{q}\partial T/T$
implies the Onsager-Kelvin relation $\Pi=TS$. For a mono-domain simple
ferroelectric all transport coefficients should be positive.

The electrocaloric properties are governed by the temperature and field dependent thermal equilibrium polarization $p_{0}\left(  E,T\right)  $ and
heat/energy $q_{0}\left(  E,T\right)  $ densities with susceptibilities $\chi
_{E}=\left(  \partial p_{0}/\partial E\right)  _{T}$ and $\chi_{T}=\left(
\partial p_{0}/\partial T\right)  _{E}.$

Metallic contacts efficiently screen the surface charges. When shorted, the
electric field in the FE vanishes except for small corrections due to a finite
screening length \cite{Mehta}. A constant applied voltage $\triangle V_{\mathrm{ext}}$ generates an electric field \(E_{\mathrm{ext}}=\triangle V_{\mathrm{ext}}/L\), but since $\partial E_{\mathrm{ext}}=0$ there can be no DC Peltier effect. A temperature difference $\triangle T_{\mathrm{ext}}$ between the contacts generates a gradient
 $\partial T_{\mathrm{ext}}=\triangle T_{\mathrm{ext}%
}/L$. The polarization current into a metal contact is dissipated
quickly without measurable effects. The principle observables are the electric
field-induced Peltier heat current and the Seebeck thermovoltage over the
contacts induced by a polarization change
\begin{equation}
\triangle V=-\int\frac{p-p_{0}}{\epsilon}dx,
\end{equation}
where $\epsilon$ is the dielectric constant.

\emph{Diffusion -} The conservation relation for the FE polarization reads%
\begin{equation}
\partial j_{p}=-\dot{p}-\frac{p-p_{0}}{\tau}%
\end{equation}
in terms of the relaxation time $\tau$. A similar equation holds for the heat
accumulation, but we assume from the outset that its relaxation is much faster
than that of the polarization, so the local temperature instantaneously adapts
to the external one. We also assume that thermalization of the non-equilibrium
ferroelectric order can be modelled by an equilibrium distribution with a local
temperature $T_{\mathrm{ext}}\left(  x,t\right)  $ and a non-equilibrium
chemical potential $\mu\left(  x,t\right)  $ \cite{Cornelissen}. It is
convenient to define $\mu_{\mathrm{tot}}=\mu-PE_{\mathrm{ext}},$ where $P$ is
the electric dipole of the unit cell such that the driving force in Eq.
(\ref{onsager}) $ E\rightarrow-\mu_{\mathrm{tot}}/P.$ With
$p-p_{0}=\chi_{E}\mu/P+\chi_{T}\left(  T-T_{0}\right)  ,$ we arrive at the
diffusion equation for the non-equilibrium chemical potential
\begin{equation}
\partial^{2}\mu-\frac{1}{\lambda^{2}}\mu=\frac{\tau}{\lambda^{2}}\left(
\dot{\mu}_{\mathrm{tot}}+\frac{P\chi_{T}}{\chi_{E}}\dot{T}_{_{\mathrm{ext}}%
}\right)
\end{equation}
with diffusion length $\lambda=\sqrt{\sigma\tau/\chi_{E}}.$ In frequency
($\omega$) space
\begin{equation}
\partial^{2}\mu-\frac{\mu}{\bar{\lambda}^{2}}=-\frac{i\omega\tau}{\lambda^{2}%
}\dot{F}_{\mathrm{ext}}\left(  x,\omega\right)
\end{equation}
where $F_{\mathrm{ext}}=\mu_{_{\mathrm{ext}}}+P\chi_{T}T_{\mathrm{ext}}%
/\chi_{E}$ and $\lambda^{2}/\bar{\lambda}^{2}=1-i\omega\tau.$ For a
capacitor with contacts at $x_{1}=0,\ x_{2}=L$, using $\partial
^{2}F_{\mathrm{ext}}=0$ and short diffusion lengths $\lambda\ll L$,
\begin{align}
\mu\left(  x,\omega\right)   &  =A\left(  \omega\right)  e^{-x/\bar{\lambda}%
}+B\left(  \omega\right)  e^{-\left(  x-L\right)  /\bar{\lambda}}\nonumber\\
&  +\frac{i\omega\tau}{1-i\omega\tau}F_{\mathrm{ext}}\left(  x,\omega\right)
, \label{Diffusion equation}%
\end{align}

\emph{Interfaces -} The boundary conditions to the contacts fix the
integration constants $A$ and $B$. The interface transport coefficients obey
an Onsager relation similar to Eq. (\ref{onsager}). Demanding continuity of
the polarization current, the boundary conditions for the chemical potential
are governed by an interface conductance $G$
\begin{equation}
G\mu\left(  0^{+}\right)  =\sigma\partial\mu\left(  0^{+}\right)
;\;G\mu\left(  L^{-}\right)  =-\sigma\partial\mu\left(  L^{-}\right)  .
\end{equation}
We focus below on two limiting cases. A good metal contact efficiently screens
the polarization dynamics and suppresses the chemical potential at the
interface. This is the transparent interface limit $G\gg\sigma$/$\lambda$. The opposite limit of an
opaque interface with $G\ll\sigma/\lambda$  represents, e.g., a contact with a 
thin non-FE spacer between FE and the metal.

\emph{Solutions -} Close to the left interface
\begin{align}
\mu\left(  x,\omega\right)   &  =\frac{\lambda}{\frac{\lambda G}{\sigma}%
+\sqrt{1-i\omega\tau}}\nonumber\\
&  +\frac{i\omega\tau}{1-i\omega\tau}\left[  \partial F_{\mathrm{ext}}\left(
\omega\right)  -\frac{G}{\sigma}F_{\mathrm{ext}}\left(  0,\omega\right)
\right]
\end{align}
In the DC limit $\omega\rightarrow0$
\begin{equation}
\mu_{\mathrm{DC}}\left(  x\right)  =\frac{\lambda}{\frac{\lambda G}{\sigma}%
+1}e^{-x/\lambda}PS\partial T_{\mathrm{ext}} \label{muDC}%
\end{equation}
The thermovoltage generated by a temperature difference
\begin{equation}
\triangle V_{\mathrm{DC}}=-\frac{\lambda}{\frac{\lambda G}{\sigma}+1}%
\frac{\chi_{E}}{\epsilon}\frac{\lambda}{L}S\triangle T_{\mathrm{ext}}%
\end{equation}
is maximized for an opaque interface $G/\sigma\rightarrow 0.$ In a symmetric
capacitor the contributions from both interfaces cancel, so the total
$\triangle V_{\mathrm{DC}}$ increases with the contrast of the interface
conductances. The theoretically possible maximal thermovoltage in the weakly
dissipative limit $\left(  L\ll\lambda\right)  $ for one transparent and one
opaque interface is $\triangle V_{\mathrm{\max}}=-\chi_{E}LS\triangle
T_{\mathrm{ext}}/\left(  2\epsilon\right)  .$

At finite frequencies the length and time scales $\lambda$ and $\tau$ govern
the dynamics. We consider the transients generated by switching on the
external perturbation $F_{\mathrm{ext}}=F_{\mathrm{ext}}^{\left(
\Theta\right)  }\Theta\left(  t\right)  ,$ where $\Theta$ is the step
function, on a time scale faster than $\tau$. The Fourier transform back to
the time domain can be obtained by contour integration.

For a transparent (left) interface we arrive at a transient accumulation in the FE caused by an electric
field pulse of
\begin{equation}
\mu\left(  x,t>0\right)  =-e^{-t/\tau} PE_{\mathrm{ext}}^{\left(
\Theta\right)  }\operatorname{erf}\left(  \frac{1}{2}\sqrt{\frac{\tau}{t}%
}\frac{x}{\lambda}\right)  .
\end{equation}
We recover the pure electrocaloric term in the bulk of the FE $\mu\left(
x\gg\lambda,t>0\right)  =-e^{-t/\tau} PE_{\mathrm{ext}}^{\left(
\Theta\right)  },$ which dominates the observable thermovoltage. The polarization current in the absence of a temperature gradient is generated by
the leakage of the electrocaloric accumulation into the contact, on the left side of the form
\begin{equation}
j_{p}\left(  x,t>0\right)  =\sigma\sqrt{\frac{\tau}{\pi t} }  
e^{- \frac{t} {\tau} -\frac{\tau}{t}\left(  \frac{x}{2\lambda}\right)  ^{2}}\frac
{E_{\mathrm{ext}}^{\left(  \Theta\right)  }}{\lambda},
\end{equation}
while that for the right contact has the opposite sign. The associated Peltier
heat current $j_{q}=\Pi j_{p}$ cools the FE and heats the contacts or
\textit{vice versa,} with a possible interface contribution.

In the opaque interface limit \begin{widetext}
\begin{align}
\mu\left(  x,t>0\right)   &  = PS\lambda\partial T_{\mathrm{ext}}^{\Theta
}\left(  e^{-x/\lambda}-\frac{1}{2}\left[  e^{-x/\lambda}%
\mathrm{erfc}\left(  \sqrt{\frac{t}{\tau}}-\frac{x}{2\lambda}\sqrt{\frac{\tau
}{t}}\right)  +e^{x/\lambda}\mathrm{erfc}\left(  \sqrt{\frac{t}{\tau}}%
+\frac{x}{2\lambda}\sqrt{\frac{\tau}{t}}\right)  \right]  \right)  \nonumber\\
&  +e^{-t/\tau}\left[  -F_{\mathrm{ext}}^{\left(  \Theta\right)  }%
+\frac{P\chi_{T}}{\chi_{E}}\partial T_{\mathrm{ext}}^{\left(  \Theta\right)
}\left(  x\operatorname{erf}\left(  \frac{1}{2}\sqrt{\frac{\tau}{t}}\frac
{x}{\lambda}\right)  +\frac{2\lambda}{\sqrt{\pi}}\sqrt{\frac{t}{\tau}%
}e^{-\frac{\tau_{E}}{t}\left(  \frac{x}{2\lambda}\right)  ^{2}}\right)
\right]
\end{align}
\end{widetext}
The polarization current vanishes at the interface $x=0$, but
\begin{align}
\mu\left(  0,t>0\right)   &  =PS\lambda\operatorname{erf}\left(  \sqrt{t/\tau
}\right)  \partial T_{\mathrm{ext}}^{\left(  \Theta\right)  }+\nonumber\\
e^{-t/\tau}  &  \left[  -F_{\mathrm{ext}}^{\left(  \Theta\right)  }\left(
0\right)  +P\frac{\chi_{T}}{\chi_{E}}\partial T_{\mathrm{ext}}^{\left(
\Theta\right)  }\frac{2\lambda}{\sqrt{\pi}}\sqrt{\frac{t}{\tau}}\right]
\end{align}
We recognize a Seebeck contribution caused by the build-up of a polarization
accumulation/depletion at the interface that approaches the DC limit Eq.
(\ref{muDC}) for long times. The second term is purely caloric but corrected by diffusion via the third term.

\emph{Phonon model -} The caloritronic parameters $\sigma$ and $\Pi$ depend on the nature of the ferroelectricity and material, but their magnitude is not known. Here we derive estimates by a simple model of one-dimensional diatomic chains at temperatures
below a displacive phase transition that generates a permanent electric dipole $P=\delta
Q$ in each unit cell, where $\delta$ is the deformation and $Q$ the ionic charge. at finite temperature the
polarization is affected by transverse phonons with maximum frequency
$\omega_{\mathrm{op}}=2\sqrt{C/M},$ where $M$ is the ionic mass and $C$ the
force constant. We adopt the ``ferron" approximation that the thermal fluctuations leave \(P\) invariant, but reduce its projection along the FE order.  For high temperatures $k_{B}T\gg
\hbar\omega_{\mathrm{op}}$ the Boltzmann equation for a constant scattering
relaxation time $\tau_{r}\ll\tau$ then leads to a Peltier coefficient
\begin{equation}
\Pi=\left.  \frac{J_{q}}{J_{p}}\right\vert _{\partial T=0}=\frac{C\delta^{2}%
}{P},
\end{equation}
and the conductivities
\begin{equation}
\sigma=\frac{\tau_{r}}{\Pi^{2}}\frac{\omega_{\mathrm{op}}^{2}}{8a}k_{B}T,
\end{equation}
\begin{equation}
\kappa=\sigma\Pi^{2}/T,
\end{equation}
where $a$ is the lattice constant. In this model, the electrocaloric properties also depend on the Peltier coefficient
\begin{equation}
\chi_{E}=\frac{k_{B}T}{a^{3}\Pi^{2}};\;\chi_{T}=\frac{k_{B}}{a^{3}\Pi}\left(
1-\frac{E}{\Pi}\right)  .
\end{equation}
A stiffer material increases the heat relative to the polarization current,
suppressing the calorics and enhancing caloritronics. The figure of merit
\begin{equation}
ZT=\left(  \sigma/\kappa\right)  \left(  \Pi^{2}/T\right)  =1
\end{equation}
does not depend on the model parameters and temperature (even when
$k_{B}T\ngtr\hbar\omega_{\mathrm{op}}$).

\textit{Estimates }- Room-temperature material constants close to
barium titanate of $\omega_{\mathrm{op}}=5\,\mathrm{THz}$, $a=0.4$\thinspace nm,
$P=2\times10^{-29}\,\mathrm{Cm},$ $\delta=0.03\,\mathrm{nm}$,
$C=25\,\mathrm{J/m}^{2}\mathrm{,}$ $\tau_{r}=1\,$ps, and $\tau=1\,$ns lead to
$\Pi=10\,\mathrm{MV/cm},$ $\chi_{E}=7\times10^{-11}\,\mathrm{C/(Vm)},$
$\chi_{T}=-2\times10^{-4}\,\mathrm{C/(Km}^{2}\mathrm{)},$ $\kappa
=4\,\mathrm{W/(Km),}$\textrm{ }$\sigma=10^{-15}\,\mathrm{m/\Omega,}$ and
$\lambda=130\,\mathrm{nm}.$ With $\epsilon/\epsilon_{0}=2000$ in terms of the
vacuum dielectric constant $\epsilon_{0}$ the predicted DC thermovoltage
induced by a temperature gradient of $\partial T=10$ $\mathrm{K/\mu m}$ at an
opaque interface is $\triangle V=2\,\mathrm{mV}$. The integrated heat flow
through a transparent interface $\int j_{p}\left(  0,t\right)  dt$ excited by
an electric field pulse of $E_{\mathrm{ext}}^{\left(  \Theta\right)
}=1\,\mathrm{MV/cm}$ is 9$\,\mathrm{J/m}^{2}.$ Because of the uncertainties in
the parameters and simplicity of the model these numbers should be taken with
a grain of salt.

\textit{Summary }- We predict caloritronic effects in planar capacitors filled
with an electrically insulating ferroelectric that may interfere with normal
operation or used for energy applications.

\textit{Acknowledgments }- The authors thank T. Teranishi, J. Kano, K. Yamamoto, and W. Yu
for valuable discussions. G.B. is supported by JSPS KAKENHI Grant No.
19H00645. R.I. and K.U. are supported by JSPS KAKENHI Grant No. 20H02609, JST
CREST \textquotedblleft Creation of Innovative Core Technologies for
Nano-enabled Thermal Management\textquotedblright\ Grant No. JPMJCR17I1, and
the Canon Foundation.

\end{document}